\documentclass[letterpaper,11pt,fleqn]{article}
 \pdfoutput=1
\usepackage{jheppub}
 
\usepackage{dcolumn}
 \usepackage{soul}
 
\usepackage{graphicx}
\usepackage{bm,amsmath,amssymb}
\usepackage[mathscr]{eucal}
 
\title{ Quasinormal modes and dispersion relations for quarkonium in a plasma }
 
 \usepackage{amsmath}

\DeclareMathOperator\Erf{Erf}
\DeclareMathOperator\Erfi{Erfi}
  
\author[a]{Nelson R. F. Braga} 
\author[a]{and Luiz F.  Ferreira }

\affiliation[a]{Instituto de F\'{\i}sica,
Universidade Federal do Rio de Janeiro, Caixa Postal 68528, RJ
21941-972 -- Brazil} 
\emailAdd{braga@if.ufrj.br}
\emailAdd{luizfaulhaber@if.ufrj.br}

\abstract{  
Recent investigations show that the thermal spectral function  of heavy $ {b \bar b } $ and 
$ {c \bar c} $ vector mesons  can be described using holography. These studies consider a bottom up model that  captures the heavy flavour spectroscopy of masses and decay constants in the vacuum  and is  consistently extended to finite temperature.  The corresponding spectral functions provide a picture of the dissociation process in terms of the decrease of the quasi-state peaks with temperature.  

Another related tool that provides important information about the thermal behaviour is the analysis of the quasinormal modes. They are field solutions in a curved background assumed to represent, in gauge/gravity duality, quasi-particle states in a thermal medium. The associated complex frequencies are related to the thermal mass and width. 
We present here the calculation of quasinormal modes for charmonium and bottomonium using the holographic approach. The temperature dependence of   mass and  thermal width are investigated.
Solutions corresponding to heavy mesons moving into the plasma are also studied. They  provide  the  dependence of the real and imaginary parts of the frequency with the quasi-particle momenta, the so called dispersion relations.

           }


\begin{document}
\maketitle

 \section{ Introduction }   
 
 The fraction of  heavy vector mesons produced in a heavy ion collision provides important information about  the possible formation of a quark gluon plasma\cite{Matsui:1986dk,Satz:2005hx}. This is so because the presence of a thermal medium  leads to the partial dissociation of these hadronic states. That is why it is important to understand the thermal behaviour of heavy mesons. In particular, the dependence of the dissociation degree on the temperature of the medium and on the state of motion of the  meson. 
 
 The thermal behaviour of heavy vector mesons can be described using  holographic bottom up models.
A holographic model for $b \bar b$ (bottomonium)  and $ c \bar c $ (charmonium), involving two dimensionfull parameters,  was proposed in Refs. 
\cite{Braga:2015jca,Braga:2016wkm,Braga:2017oqw}. An improved model containing three energy parameters  appeared then in Refs. \cite{Braga:2017bml,Braga:2018zlu}. These parameters have a simple physical interpretation. They represent:  the quark mass and the string tension, that are related to the mass spectra,  and an  ultraviolet (UV) energy scale, necessary in order  to fit the  decay constant spectra.  This UV energy parameter is related to the large mass change that occurs in a non hadronic decay, when a heavy vector meson  transforms into  light leptons. 

The thermal spectral function  for heavy vector mesons was constructed using this holographic approach 
in references \cite{Braga:2017bml,Braga:2018zlu}. Quasi-Particle states appear as peaks that broadens when the temperature increases. 

Important information about the behaviour of hadrons inside a thermal medium can be obtained also from the so called quasinormal modes. They are gravitational field solutions that play the role of gravity duals to quasi-particle states. The associated frequencies are complex and the imaginary parts are related to the thermal width of the quasi-state. Quasinormal modes are the finite temperature version of the normalized solutions, that describe particle states at zero temperature in holography.

We will develop here the calculation of quasinormal modes using the bottom up holographic model of references \cite{Braga:2017bml,Braga:2018zlu}.  The dependence of the real and imaginary part of the frequency on the temperature will be investigated for charmonium and bottomonium states. Then the dispersion relations will be considered. The dependence of the complex frequencies on the linear momentum, for hadrons moving inside a plasma, will be analysed for different temperatures. 
This type of analysis provides a detailed picture of the thermal behaviour of the heavy vector mesons in
a medium like the quark gluon plasma. 

The article is organized in the following way. In section { \bf 2} we review the holographic model for charmonium and bottomonium. Then in  section {\bf 3} we study the spectral functions for charmonium and bottomonium. The quasinormal modes are calculated  in section {\bf  4} and  conclusions and final comments are shown in section {\bf 5}.

\section{Holographic model for heavy vector mesons }

The  bottom up holographic models for heavy vector mesons of refs. \cite{Braga:2017bml,Braga:2018zlu} 
are defined, in the zero temperature case, in 5-d anti-de Sitter space-time, with metric 
\begin{equation}
 ds^2 \,\,= \,\, \frac{R^2}{z^2 }(-dt^2 + d\vec{x}\cdot d\vec{x} + dz^2)\,.
 \label{AdSmetric}
\end{equation}
The gravity duals of heavy vector mesons are  fields $V_m = (V_\mu,V_z)\,$ ($\mu = 0,1,2,3$), that are assumed to represent  the gauge theory currents $ J^\mu = \bar{\psi}\gamma^\mu \psi \,$. The action integral is:
\begin{equation}
I \,=\, \int d^4x dz \, \sqrt{-g} \,\, e^{- \phi (z)  } \, \left\{  - \frac{1}{4 g_5^2} F_{mn} F^{mn}
\,  \right\} \,\,, 
\label{vectorfieldaction}
\end{equation}
where $F_{mn} = \partial_m V_n - \partial_n V_m$. The energy parameters are introduced through the background scalar field $\phi(z)$.   The scalar field used in 
ref.\cite{Braga:2017bml}  
\begin{equation}
\phi(z)=k^2z^2 + \tanh\left(\frac{1}{Mz}-\frac{k}{ \sqrt{\Gamma}}\right)\,,
\label{dilatonMod0}
\end{equation} 
was usefull  for charmonium states but did not provide a nice phenomenological description of the spectra of decay constants and masses for bottomonium. The improved version, presented in ref.  
\cite{Braga:2018zlu}, that we will use here,  provides a nice fit for both chamonium and bottomonium states. It has the following form:  
\begin{equation}
\phi(z)=k^2z^2+Mz+\tanh\left(\frac{1}{Mz}-\frac{k}{ \sqrt{\Gamma}}\right)\,.
\label{dilatonModi}
\end{equation} 

  The parameters $k$ and $\Gamma $ have simple interpretations in terms of just the mass spectra of mesons made up of two heavy quarks:  $k$ is related to the  quark mass while  $\Gamma $, with dimension of energy squared, is related to the string tension of the strong quark anti-quark interaction. On the other hand, the third parameter $M$ has a more subtle interpretation.
Heavy vector mesons undergo non hadronic decay processes, when the final state consists of light leptons. In such transitions, that are not governed by the strong interaction, there is a a very large mass  change. The parameter $ M$ is  representing effectively the mass scale of such a transition, characterized by a matrix element of the form   $ \langle 0 \vert \, J_\mu (0)  \,  \vert n \rangle = \epsilon_\mu f_n m_n \,, $
representing a transition from a meson  at radial excitation level $n$ to the hadronic vacuum. 
 The decay constant $ f_n $ is essentially proportional to this matrix element and  the large mass parameter $M$ makes it possible to fit the corresponding spectra.

 The simplest way to realise gauge gravity duality is to gauge away the $z$ component: $V_z=0$. 
 Then the remaining $V_\mu $ work as sources for the currents $ J^\mu $.  The equation of motion for the $ \mu = (1,2,3)$ components, denoted here generically as $V$, in momentum space reads
\begin{equation}
\partial_{z} \left[  \frac{R}{z} e^{-\phi(z)} \partial_{z} V (p,z)  \right]-p^2  \frac{R}{z} e^{-\phi(z)} V (p,z)  = 0\,.
\label{eqmotion}
\end{equation}
 
The normalizability requirement for  solutions of equation of motion (\ref{eqmotion})  corresponds to the 
boundary condition  $ V (p, z=0) =0$. The corresponding solutions $ V(p,z)=\Psi_n(z)$ show up for a discrete spectrum of $p^2=-m_{n}^{2}$ where $m_n$ are interpreted as  the masses of the corresponding meson states.

Decay constants are  proportional to the transition matrix from  the vector meson $n$ excited state to the vacuum:  $  \langle 0 \vert \, J_\mu (0)  \,  \vert n \rangle = \epsilon_\mu f_n m_n $. 
In the present  holographic model they are given by\cite{Braga:2017bml} 
\begin{equation}
f_n=\frac{1}{g_{5} m_{n}}\lim\limits_{z \rightarrow 0} \left( \frac{R}{z} e^{-\phi(z)} \partial_z  \Psi_n(z)\right) \,.
\label{decayconstant}
\end{equation}

 The  values of the parameters that  describe  charmonium and bottomonium are respectively:
\begin{eqnarray}
  k_c  = 1.2  \, {\rm GeV } ; \,\,   \sqrt{\Gamma_c } = 0.55  \, {\rm GeV } ; \,\, M_c=2.2  \, {\rm GeV }\,;
  \label{parameters1}
  \\
   k_b = 2.45  \, {\rm GeV } ; \,\,   \sqrt{\Gamma_b } = 1.55  \, {\rm GeV } ; \,\, M_b=6.2  \, {\rm GeV }\,.
  \label{parameters2}
  \end{eqnarray}  
\begin{table}[h]
\centering
\begin{tabular}[c]{|c||c||c|}
\hline 
\multicolumn{3}{|c|}{  Holographic (and experimental) results for bottomonium   } \\
\hline
 State &  Mass (MeV)     &   Decay constants (MeV) \\
\hline
$\,\,\,\, 1S \,\,\,\,$ & $ 6905 \,\,(9460.30\pm 0.26) $  & $ 719 \,( 715.0 \pm 2.4) $ \\
\hline
$\,\,\,\, 2S \,\,\,\,$ & $   8871 \,( 10023.26 \pm 0.32) $   & $ 521 \,(497.4 \pm 2.2) $  \\
\hline 
$\,\,\,\,3S \,\,\,\,$ & $  10442 \, \,( 10355.2 \pm 0.5) $   & $427 \, (430.1  \pm 1.9) $ \\ 
\hline
$ \,\,\,\, 4S  \,\,\,\,$ & $ 11772 \, (10579.4 \pm 1.2)  $  & $ 375 \,(340.7  \pm 9.1)$ \\
\hline
\end{tabular}   
\caption{Holographic masses and decay constants for the bottomonium $S$-wave resonances. Experimental values inside parentheses for comparison. }
\end{table}

   \begin{table}[h]
\centering
\begin{tabular}[c]{|c||c||c|}
\hline 
\multicolumn{3}{|c|}{  Holographic (and experimental)  results for charmonium   } \\
\hline
 State &  Mass (MeV)     &   Decay constants (MeV) \\
\hline
$\,\,\,\, 1$S$ \,\,\,\,$ & $ 2943 \,\, (3096.916\pm 0.011)  $  & $ 399 \, (416 \pm 5.3)$ \\
\hline
$\,\,\,\, 2$S$ \,\,\,\,$ & $  3959 \,\, (3686.109 \pm 0.012) $   & $ 255  \, (296.1 \pm 2.5)$  \\
\hline 
$\,\,\,\,3$S$ \,\,\,\,$ & $  4757 \,\, (4039 \pm 1 ) $   & $198 \, ( 187.1  \pm 7.6) $ \\ 
\hline
$ \,\,\,\, 4$S$  \,\,\,\,$ & $ 5426\,\,  (4421 \pm 4)  $  & $ 169 \,  (160.8  \pm 9.7)$ \\
\hline
\end{tabular}   
\caption{Holographic masses and decay constants for the charmonium $S$-wave resonances. Experimental values inside parentheses for comparison.  }
\end{table}

We  show on Tables 1 and 2 the results for charmonium and bottomonium, calculated in ref.  \cite{Braga:2018zlu}  using  this model.
For comparison we shown inside parentheses the corresponding available experimental  data.  
The masses come directly from \cite{Agashe:2014kda} while decay constants are obtained from  the experimental values of masses and electron positron decay widths $\Gamma_{ V \to e^+ e^ - } $ using the relation \cite{Hwang:1997ie}:
\begin{equation} 
f_{_V}^ 2 \,=\, \frac{3 m_{_V} \Gamma_{ V \to e^+ e^ - }  }{4 \pi \alpha^ 2 c_{_V}}\,,
 \label{decay-widths}
 \end{equation} 
  \noindent where $\alpha = 1/137 $    and the charge coefficients $c_{_V} $ for the two families that we consider are  $  c_{c \bar c}  = 4/9 $ and $c_{   b \bar b} = 1/9$. 
  Decay constants obtained from experimental data decrease monotonically with the radial excitation level for both charmonium and bottomonium states. 
   It  is important to note the very nice fit of the decay constants provided by the model compared with those obtained from experimental data. 
 This was the main motivation for the construction of  this alternative approach to the holographic description of heavy vector mesons.   
Holographic models provide in general results that are not compatible with the experimental data regarding decay constants . For example, the hard wall model, 
 proposed in refs. \cite{Polchinski:2001tt,BoschiFilho:2002ta,BoschiFilho:2002vd}, provides decay constants that increase with the radial excitation number, while the soft wall model \cite{Karch:2006pv} leads to decay constants that are degenerate with respect to the radial excitation level.  

 Now, one could ask: why is it important to have a nice  for the zero temperature decay constants when one aims to find an  appropriate description of the finite temperature thermal behavior?  It is simple to understand this fact, once we remind us about the 
connection   between decay constants and the spectral function.  

The thermal spectral function is the imaginary part of the retarded Green's function. 
The relevant part of the Green's function is the two point function that, at zero temperature, has a spectral decomposition  in terms of masses $m_n$ and   decay constants $f_n$  of the states:  
 \begin{equation}
\Pi (p^2)  = \sum_{n=1}^\infty \, \frac{f_n^ 2}{(- p^ 2) - m_n^ 2 + i \epsilon} \,.
\label{2point}
\end{equation}
The imaginary part of eq.(\ref{2point}) is a sum of delta peaks with coefficients proportional to the square of the decay constants: $ f_n^2 \, \delta ( - p^2 - m_n^2 ) $. 
At finite temperature, the quasi-particle states appear in the spectral function as broader  peaks  as the temperature $T$   of the medium increase. This analysis strongly suggests that a consistent  extension of a hadronic model to finite temperature should take into account the zero temperature behavior, where decay constants play an essential role.  
    
   The extension to finite temperature is obtained by replacing the AdS space of eq. (\ref{AdSmetric}) by  an AdS  black hole  geometry 
\begin{equation}
 ds^2 \,\,= \,\, \frac{R^2}{z^2}  \,  \Big(  -  f(z) dt^2 + d\vec{x}\cdot d\vec{x}  + \frac{dz^2}{f(z) }    \Big)   \,,
 \label{metric2}
\end{equation}
with $ f (z) = 1 - \frac{z^ 4}{z_h^4} $, where $z_h$ is the horizon position. The black hole temperature comes from the requirement of absence of a conical singularity at the horizon, in the Euclidean version of the metric. 
In this imaginary time formulation, the time variable is periodic, with period $ 0 \le t \le \beta = 1/T$, where $T$ is the temperature. 
 The regularity of the metric leads to 
\begin{equation} 
T =  \frac{\vert  f'(z)\vert_{(z=z_h)}}{4 \pi  } = \frac{1}{\pi z_h}\,.
\label{temp}
\end{equation}


\section{Thermal spectral function    }

\subsection{Equations of motion} 
The equations of motion for the holographic model come from action (\ref{vectorfieldaction}) with the metric (\ref{metric2}).  One chooses again the radial gauge $V_{z}=0$ and  consider plane wave solutions of the form: $V_{\mu}(z,x_{1},x_{2},x_{3},t)=e^{-\omega t+i q x_{3}}V_{\mu}(z,\omega,q)$ propagating in the $x_{3}$  direction with the wave vector $p_{\mu}=(-\omega,0,0,q)$. The equations have the following form
\begin{equation}\label{eqtt}
V_{t}''-\left(\frac{1}{z}+\phi' \right)V_{t}'-\frac{q}{f}\left(qV_{t}+\omega V_{3}\right)=0,
\end{equation} 
\begin{equation}\label{eqx1x1}
V_{\beta}''+\left(\frac{f'}{f}-\frac{1}{z}-\phi' \right)V_{\beta}'+\frac{1}{f^2}\left(\omega^2-q^2f \right)V_{\beta}=0 \, , \,\,\,\,\,\,\,\,\,\,\ (\beta=1,2)
\end{equation}

\begin{equation}\label{eqx3x3}
V_{3}''+\left(\frac{f'}{f}-\frac{1}{z}-\phi' \right)V_{3}'+\frac{\omega}{f^2}\left(qV_{t}+\omega V_{3}\right)=0, \,\,\,\,\,\,\,\,\,\,\ 
\end{equation} 

\begin{equation}\label{eqzz}
\omega V_{t}'+qfV_{3}'=0
\end{equation}

\noindent where  the  prime  ($'$)  denotes  the  derivative  with  respect  to z. The corresponding equations in terms of the electric field components: $E_{1}=\omega V_{1}$, $E_{2}=\omega V_{2}$ and $E_{3}=\omega V_{3}+qV_{t}$, are given by

\begin{equation}\label{eqTrans}
E_{\alpha}''+\left(\frac{f'}{f}-\frac{1}{z}-\phi' \right)E_{\alpha}'+\frac{\omega^2-q^2f}{f^2}E_{\alpha}=0 \, , \,\,\,\,\,\,\,\,\,\,\ (\alpha=1,2)
\end{equation} 

\begin{equation}\label{eqLog}
E_{3}''+\left(\frac{\omega^2}{\omega^2-q^2f}\frac{f'}{f}-\frac{1}{z}-\phi' \right)E_{3}'+\frac{\omega^2-q^2f}{f^2}E_{3}=0. \,\,\,\,\,\,\,\,\,\,\ 
\end{equation}

\subsection{  Retarded Green's function }

In the four dimensional vector gauge theory we define a retarded Green's functions of the  currents  $ J_{\nu} $ as 
\begin{equation}\label{Green}
G^{R}_{\mu \nu}(p)=-i\int d^4 x e^{- i p \cdot x}\theta(t)\langle \left[ J_{\mu}(x),J_{\nu}(0)\right]\rangle\,.
\end{equation}
Current conservation implies that $ p^\mu G^{R}_{\mu \nu}(p) = 0$. 
So, the structure of the Greens function at zero temperature can be written in terms of a projector that makes explicit this property:
\begin{equation}\label{GreenProjector}
G^{R}_{\mu \nu}(p)= P_{\mu \nu}  \, \Pi ( p^2)  \,,
\end{equation}
where
\begin{equation} 
 P_{\mu \nu} = \eta_{\mu \nu}-\frac{p_{\mu}p_{\nu}}{p^2}\,.
\end{equation}
At finite temperature, in thermal equilibrium, it is interesting to separate this projector  into transverse and longitudinal parts \cite{Kovtun:2005ev} introducing 
\begin{eqnarray}\label{Projectors}
P_{00}^{T} = 0\,, \,\,\,\, P^{T}_{0i}&=& 0, \,\,\,\,\, P^{T}_{ij}=\delta_{ij}-\frac{p_{i}p_{j}}{\textbf{p}^2} \cr
 P^{L}_{\mu \nu}&=&P_{\mu \nu}-P_{\mu \nu}^{T}  .
\end{eqnarray}
Then the retarded Green's function can be writen in the finite temperature case when there is rotation invariance as   
\begin{equation}\label{Green2}
G^{R}_{\mu \nu}(p)=P^{T}_{\mu \nu}\Pi^{T}(p_{0},\textbf{p}^2)+P^{L}_{\mu\nu}\Pi^{L}(p_{0},\textbf{p}^2) \,,
\end{equation}
where $\Pi^{T}(p_{0},\textbf{p}^2)$ and $\Pi^{L}(p_{0},\textbf{p}^2)$ are independent scalar functions. 
 
Choosing,  as in the previous section, the wave vector with the form $p_{\mu}=(-\omega,0,0,q)$, corresponding to  propagation in the $z$ direction, the relevant (non-vanishing)  components of 
the Green's function can all be written in terms of the longitudinal and transversal scalar functions 
 \begin{equation}\label{Greenxx}
G_{11}^{R}(p)=G^{R}_{22}=\Pi^{T}(\omega,q)\,,\,\,\,\,\,\, G_{33}^{R}(p)=\frac{\omega^2}{\omega^2-q^2}\Pi^{L}(\omega,q) \, ,
\end{equation}
\begin{equation}\label{Greentt}
G_{tt}^{R}(p)=\frac{q^2}{\omega^2-q^2}\Pi^{L}(\omega,q)\,,\,\,\,\,G_{t3}^{R}(p)=-\frac{q\omega}{\omega^2-q^2}\Pi^{L}(\omega,q)\,.
\end{equation}

Considering now  the holographic  approach, the gauge theory current correlators 
are  represented in terms of the vector fields living in the five dimensional space and described by the action integral of eq. (\ref{vectorfieldaction}) with metric (\ref{metric2}).  In momentum space the on shell action takes the form
\begin{equation}\label{actionelec}
S=\frac{R}{2g^2_{5}}\int\frac{d\omega dq}{(2\pi)^2}\left[\frac{e^{-\phi}}{z}\Big\{V_{t}(z,-p)\partial_{z}V_{t}(z,p)-f\textbf{V}(z,-p)\cdot\partial_{z}\textbf{V}(z,p)\Big\}\right]^{z_h}_{0}  \,.
\end{equation}
 
It is convenient to express this action in terms of the electric field components
\begin{equation}\label{actionelec2}
S=-\frac{R}{2g^2_{5}}\int\frac{d\omega dq}{(2\pi)^2}\left[ \frac{e^{-\phi}}{z} \frac{f}{\omega^2}\sum_{j=1}^3\left(1-\frac{q^2}{\omega^2}f\right)^{-\delta_{j3}}E_{j}(z,-p)\partial_{z}E_{j}(z,p)\right] \,.
\end{equation}
The $z $ dependent part of the field, the so called bulk to boundary propagator,  can be separated from the boundary value   $E^0_{j}$ as
\begin{equation}\label{elecfield}
E^{(-)}_{j}(z,p)=\mathcal{E}_{j}(z,p)E^{0}_{j}(p)\, , \,\,\,\,\,\, (j=1,2,3)
\end{equation}
where the functions $\mathcal{E}_{j}(z,p)$ are defined to satisfy $\mathcal{E}_{j}(0,p)=1$ and the superscript $(-)$ indicates that $E^{(-)}_{j}(z,p)$ satisfies the infalling condition at the horizon, as required by the  Lorentzian form of Son-Starinets prescription \cite{Son:2002sd}. Then, substituting (\ref{elecfield}) in the action (\ref{actionelec2}) one finds
\begin{equation}
S=-\frac{R}{2g^2_{5}}\int\frac{d\omega dq}{(2\pi)^2}\left[ \frac{e^{-\phi}}{z} \frac{f}{\omega^2}\sum_{j=1}^3\left(1-\frac{q^2}{\omega^2}f\right)^{-\delta_{j3}}E_{j}^{0}(-p)\mathcal{E}_{j}(z,p)\partial_{z}\mathcal{E}_{j}(z,p)E^{0}_{j}(p)\right]^{z_h}_{0}
\end{equation}
In terms of the boundary values of the potential this action reads
\begin{eqnarray}
S&=&\frac{R}{2g^2_{5}}\int \frac{d\omega dq}{(2\pi)^2}\Bigg[ \frac{e^{-\phi}}{z}f\Bigg( \frac{\omega^2}{\omega^2-q^2f}\Big\{V^{0}_{3}(-p)V^{0}_{3}(p)+\frac{q}{\omega}V^{0}_{3}(-p)V^{0}_{t}(-p)-\frac{q}{\omega}V^{0}_{t}(-p)V^{0}_{3}(-p) \cr &+&\frac{q^2}{\omega^2}V_{t}^{0}(-p)V_{t}^{0}(p) \Big\}\mathcal{E}_{3}(z,-p)\partial_{z}\mathcal{E}_{3}(z,p)+\sum_{\alpha=1}^2 V^{0}_{\alpha}(-p)V^{0}_{\alpha}(p)\mathcal{E}_{\alpha}(z,-p)\partial_{z}\mathcal{E}_{\alpha}(z,p) \Bigg) \Bigg]^{z_h}_{0}
\end{eqnarray}
Finally, using the Son-Starinets \cite{Son:2002sd} prescription for the vector field case one finds 
\begin{eqnarray}\label{Greenrel}
&&\frac{G^{R}_{tt}}{q^2}=-\frac{G^{R}_{tx_{3}}}{q\omega}=-\frac{G^{R}_{3t}}{q\omega}=\frac{G^{R}_{33}}{\omega^2}=-\frac{R}{g^2_{5}}\frac{1}{\omega^2-q^2}\lim_{z \rightarrow 0} \frac{fe^{-\phi}}{z}\partial_{z}\mathcal{E}_{3}(z,p)\cr \ && G^{R}_{11}=-\frac{R}{g^2_{5}}\lim_{z \rightarrow 0} \frac{fe^{-\phi}}{z}\partial_{z}\mathcal{E}_{1}(z,p)\,, \,\,\, G^{R}_{22}=-\frac{R}{g^2_{5}}\lim_{z \rightarrow 0} \frac{fe^{-\phi}}{z}\partial_{z}\mathcal{E}_{{2}}(z,p)\,.
\end{eqnarray}

\subsection{Spectral Function}

In order to calculate the spectral function it is convenient write the  equations of motion for the electric fields (\ref{eqTrans}) and (\ref{eqLog}) in terms of the bulk to boundary propagator $E_{j}(z,p)=\mathcal{E}_{j}(z,p)E^{0}_{j}(p)$, 
\begin{equation}\label{eqTransx}
\partial_{z}^{2}\mathcal{E}_{\alpha}+\left(\frac{\partial_{z}f}{f}-\frac{1}{z}-\partial_{z}\phi \right)\partial_{z}\mathcal{E}_{\alpha}+\frac{\omega^2-q^2f}{f^2}\mathcal{E}_{\alpha}=0 \, , \,\,\,\,\,\,\,\,\,\,\ (\alpha=1,2)
\end{equation} 

\begin{equation}\label{eqLogx}
\partial_{z}^{2}\mathcal{E}_{3}+\left(\frac{\omega^2}{\omega^2-q^2f}\frac{\partial_z f}{f}-\frac{1}{z}-\partial_{z}\phi \right)\partial_{z}\mathcal{E}+\frac{\omega^2-q^2f}{f^2}\mathcal{E}_{3}=0, \,\,\,\,\,\,\,\,\,\,\ 
\end{equation}
Now the spectral function can be extracted using the relation (\ref{Greenrel}) presented in the last section. Particularly, in the case of $G^{R}_{x_3 x_3}$ and $G^{R}_{\alpha \alpha}$, the corresponding spectral functions in terms of the function $\mathcal{E}_{j}$ are
\begin{equation}\label{eq1}
\rho_{33}(\omega,q)\equiv -2 ImG^{R}_{33}(\omega,q)=\frac{2R}{g^2_{5}}\frac{\omega^2}{\omega^2-q^2}\lim_{z \rightarrow 0} \frac{fe^{-\phi}}{z}\partial_{z}\mathcal{E}_{{3}}(z,p)\,,
\end{equation} 
\begin{equation}\label{eq2}
\rho_{\alpha \alpha}(\omega,q)\equiv -2 ImG^{R}_{\alpha \alpha}(\omega,q)=-\frac{2R}{g^2_{5}}\lim_{z \rightarrow 0} \frac{fe^{-\phi}}{z}\partial_{z}\mathcal{E}_{\alpha}(z,p)\,,
\end{equation}
where $\mathcal{E}$ is the solution of the eqs. (\ref{eqTransx}) and (\ref{eqLogx}) satisfying the infalling  boundary condition
\begin{eqnarray}\label{boundary}
&& \mathcal{E}(z \rightarrow z_h,\omega) \longrightarrow \left(1-\frac{z}{z_h}\right)^{-i\omega/4\pi T}\,\left[1+a_1\left(1-\frac{z}{z_h}\right)+a_2\left(1-\frac{z}{z_h}\right)^2+...\right]\,,
\end{eqnarray}
and the bulk to boundary condition 
\begin{equation}
\label{boundary2}
\mathcal{E}(z\to 0,\omega)=1\,.
\end{equation}

\subsection{Numerical Results}
\begin{figure}[h]
\label{g67}
\begin{center}
\includegraphics[scale=0.8]{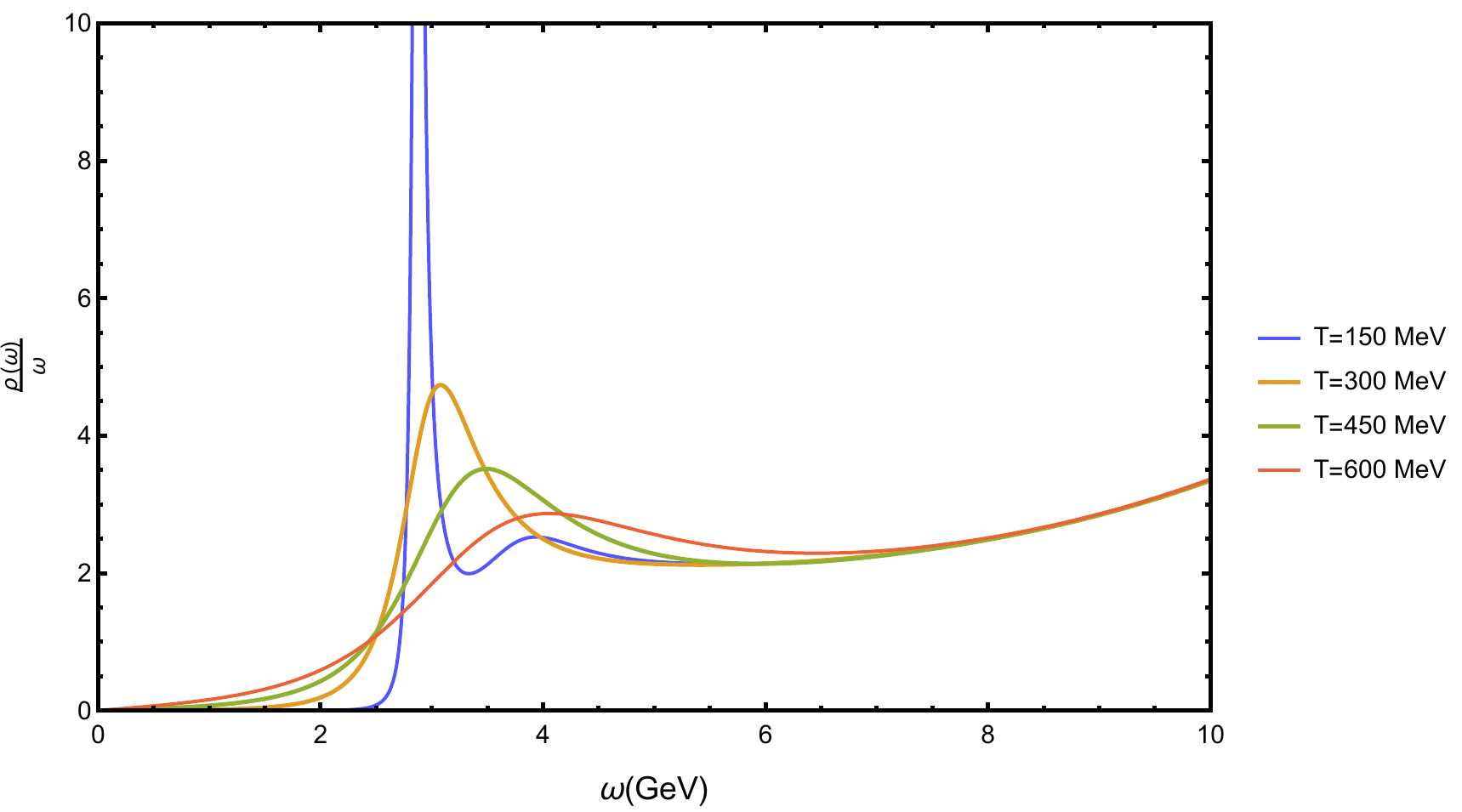}
\end{center}
\caption{Spectral function for charmonium at rest for different values of  temperature.}
\end{figure}
We performed the numerical analysis of the spectral functions for the transverse and longitudinal sectors solving the equations (\ref{eqTransx}) and (\ref{eqLogx}),  for the charmonium and bottomonium cases,  using the boundary conditions (\ref{boundary}),(\ref{boundary2}) for the fields. Then, relations (\ref{eq1}) and (\ref{eq2}) were used to find the spectral functions. The model parameters are the zero temperature ones,  presented in section {\bf 2}.  
\begin{figure}[h]
\label{g67}
\begin{center}
\includegraphics[scale=0.8]{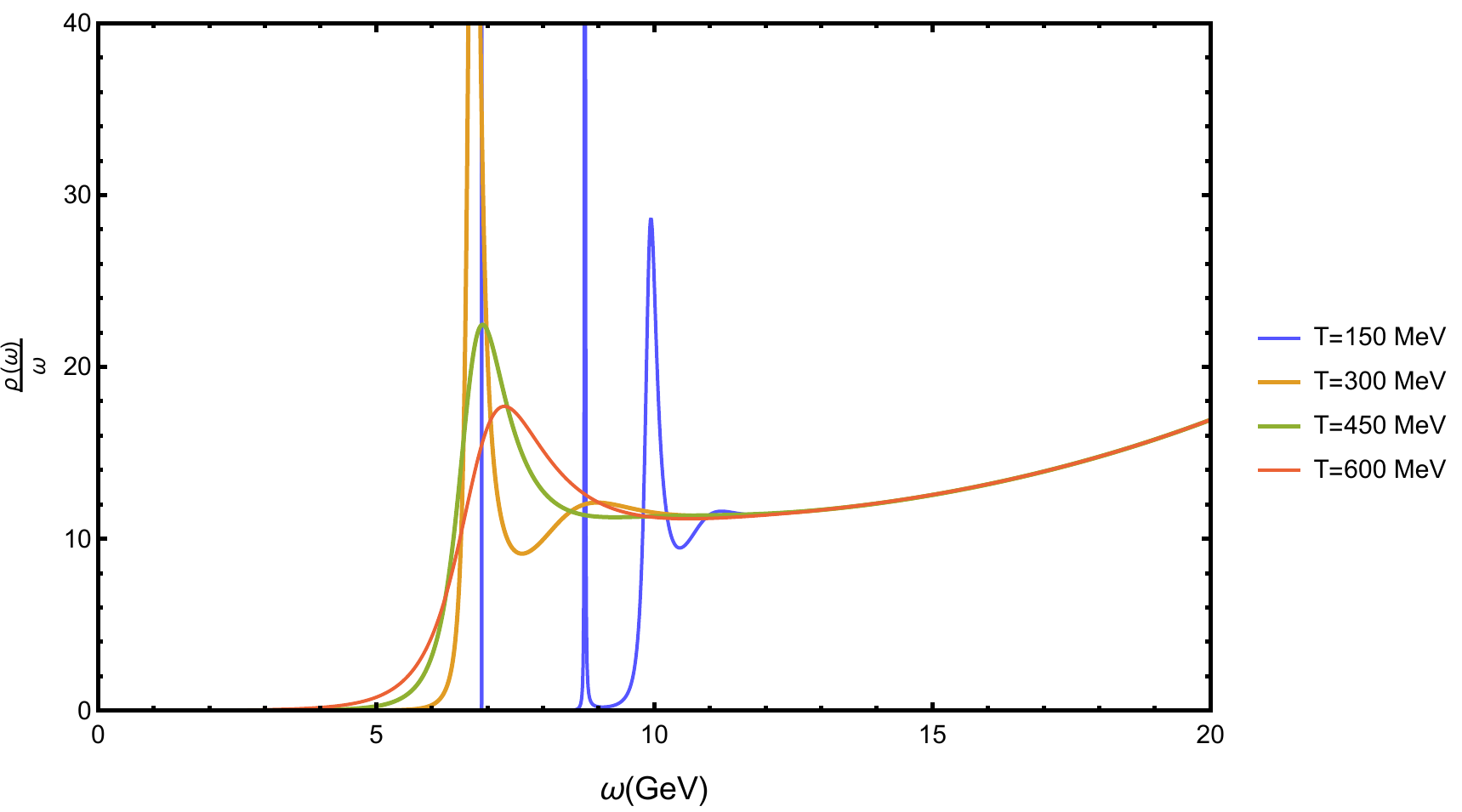}
\end{center}
\caption{Spectral function for bottomonium at rest for different values of  temperature.}
\end{figure}

\begin{figure}[h]
\label{g67}
\begin{center}
\includegraphics[scale=0.5]{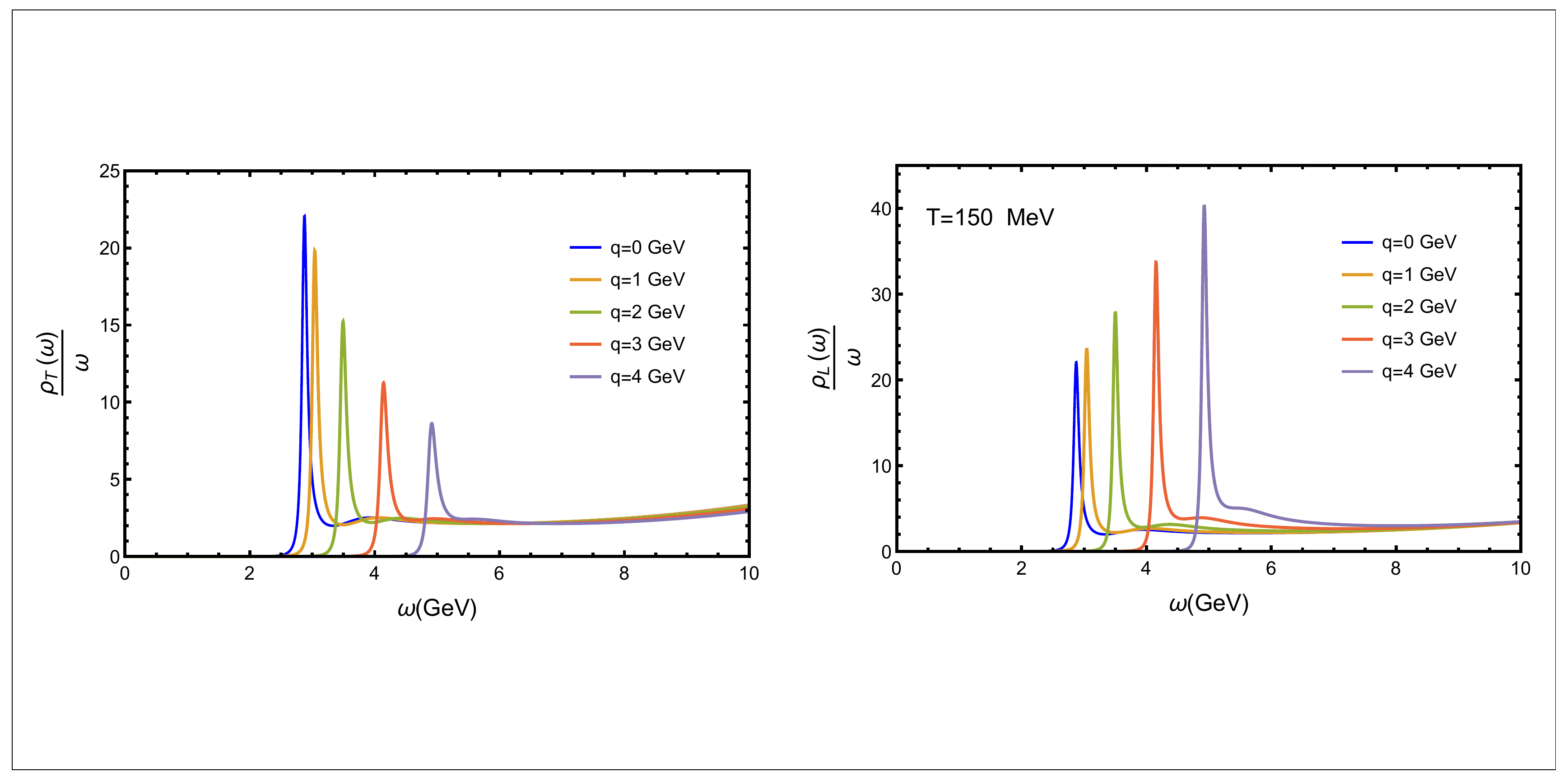}
\end{center}
\caption{Spectral Function  for  charmonium with different values of linear momentum at fixed value of temperature. }
\end{figure}

\begin{figure}[h]
\label{g67}
\begin{center}
\includegraphics[scale=0.45]{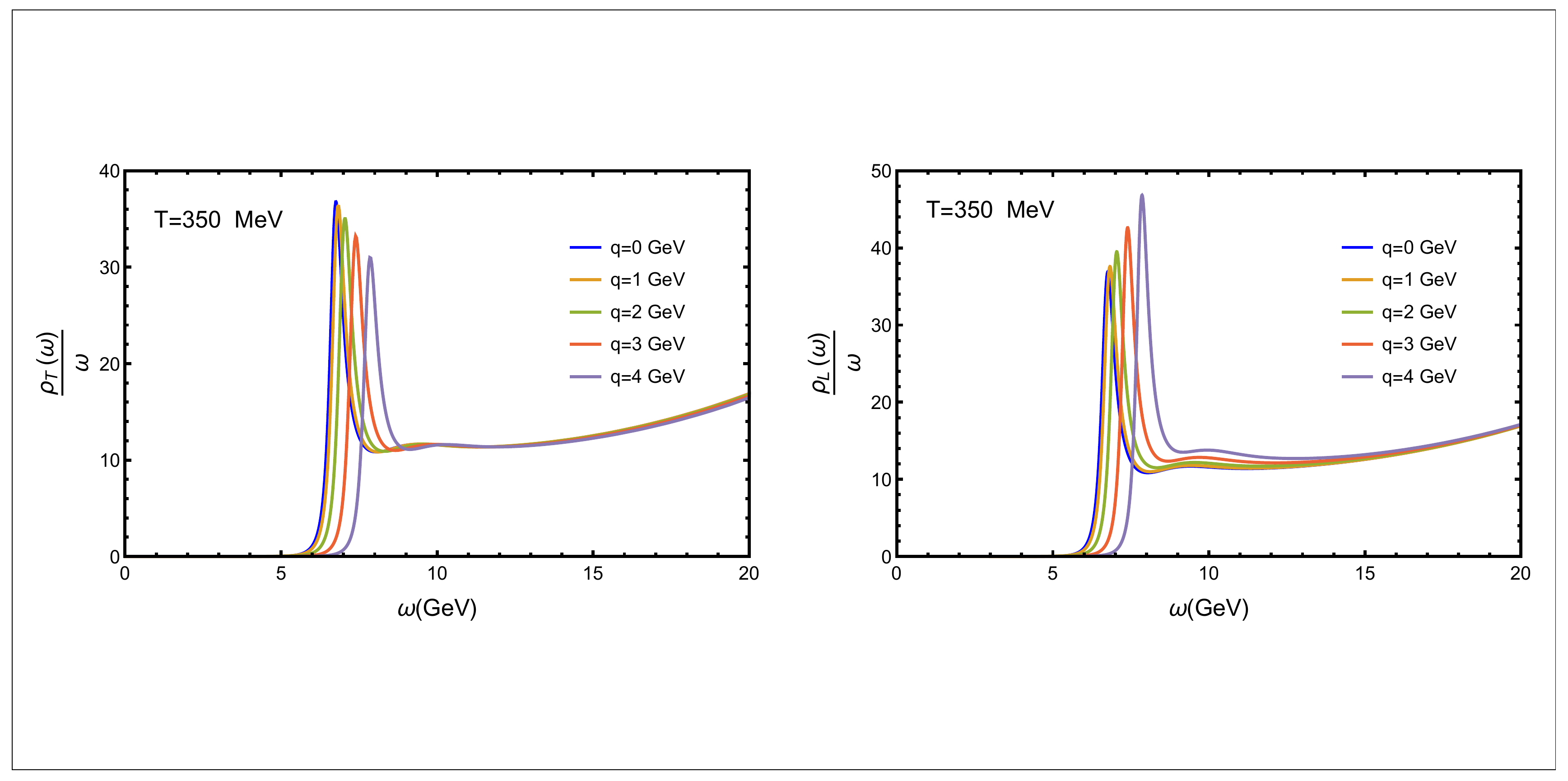}
\end{center}
\caption{Spectral Function  for  bottomonium with different values of linear momentum at fixed value of temperature. }
\end{figure}

Let us start with the spectral functions for the vector mesons at rest. In this case the spectral function is the same for the transverse and longitudinal directions. In  figure 1 we show the dissociation process at finite temperature for charmonium and in figure 2  the  bottomonium case. Note that for both flavours the peaks broadens as the temperature increases. In particular, the peak broadens  faster for charmonium than for bottomonium.

We also studied the spectral function at non vanishing momentum. In figure 3 and 4 the results obtained  for charmonium and bottomonium, respectively, are shown at different value of momentum. In both cases one can observe  the broadening that  the height  of the peaks in the longitudinal direction increase as the momentum increases while, for the transverse direction,  the height of the peaks  decrease. This can be interpreted as meaning that the quasiparticle states in longitudinal   motion are more stable in the longitudinal than  at rest ($q=0$). On the other hand,  quasiparticles
in transverse motion are less stable than at rest. The localization of the peaks in terms of the frequencies  change in both cases.


\section{ Quasinormal modes}

The peaks shown in figure 1  and 2 indicate that the corresponding  retarded Green's functions present  poles. These poles are related to the frequencies of the electromagnetic quasinormal modes of the black brane. The quasinormal modes correspond, in the dual gauge theory, to quasi-particle states of vector mesons. The frequencies of these quasinormal modes present real  and imaginary  parts, $\omega=\omega_{R}  + i \omega_{I}$.  The real part is related to the mass of the vector mesons when $q=0$, while the imaginary part to the decay rate of the quasi-particle states formed near the confining/deconfining transition. One observes that when the temperature increases, the widths of the peaks increase and the mean life $\tau=2\pi/ \omega_{I}$ decrease.

\begin{figure}[h]
\label{g67}
\begin{center}
\includegraphics[scale=0.5]{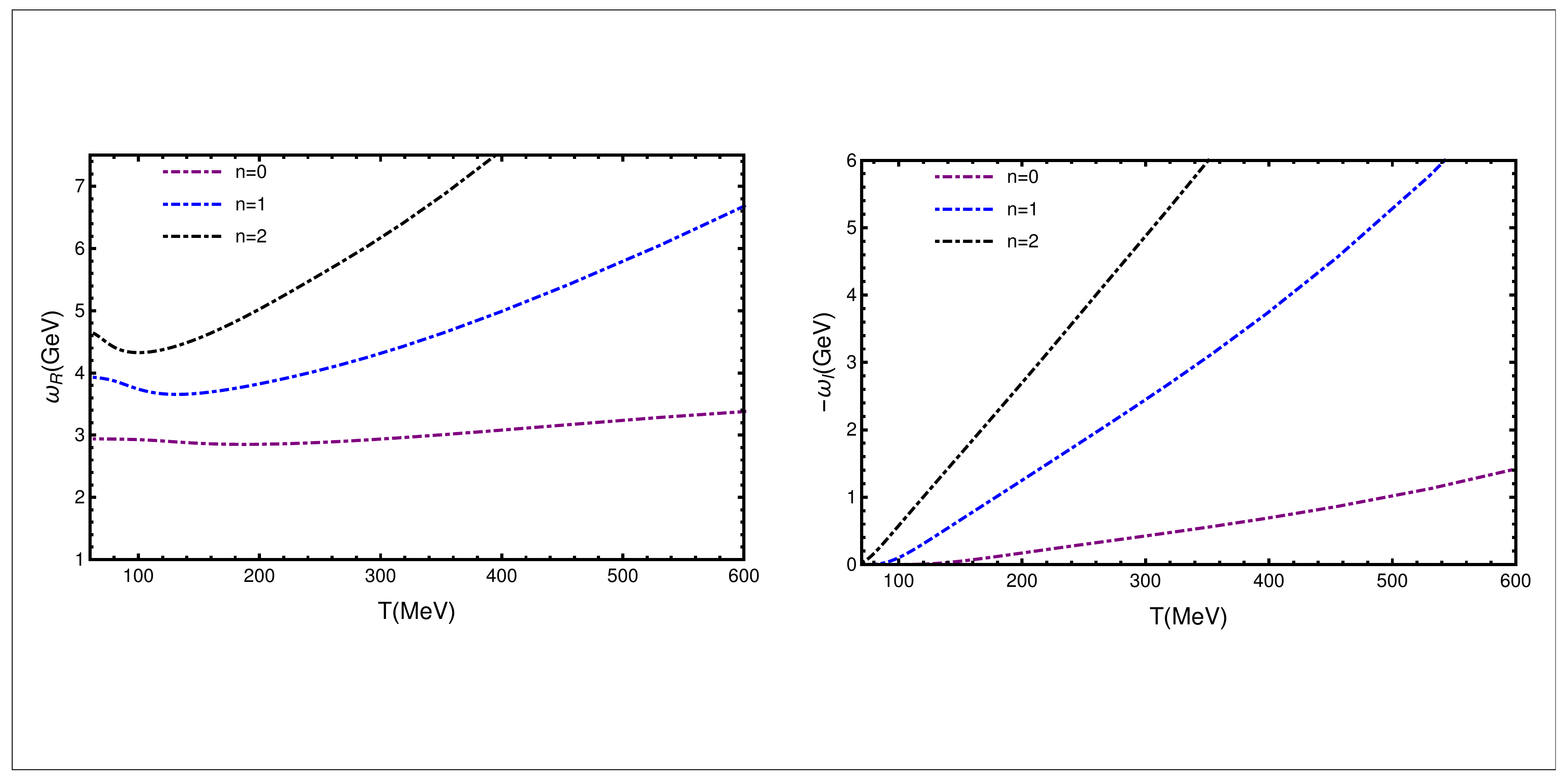}
\end{center}
\caption{Real and imaginary parts of the  frequencies of charmonium quasinormal modes as function of the temperature. The real part of the frequencies is shown on the left side, and the imaginary part on the right side. }
\end{figure}

In contrast to the zero temperature case, where there are particle states, represented by normalized solutions, in the finite temperature case we have quasi-particles, described by quasinormal modes.
They are field solutions in the curved background, subject infalling condition at the horizon.

Previous studies of quasinormal modes in the context of gauge/gravity duality
can be found, for example, in Refs. \cite{Horowitz:1999jd,Son:2002sd,Nunez:2003eq,Kovtun:2005ev,Miranda:2005qx,Hoyos:2006gb,Berti:2009kk,Morgan:2009vg,Miranda:2009uw,Konoplya:2011qq,Mamani:2013ssa,Mamani:2018qzl}.  It is interesting to note that when one considers the conformal holographic AdS/CFT case, like in refs. \cite{Son:2002sd,Kovtun:2005ev}, there are no dimensionfull fixed parameters in the theory. So, at finite temperature, the only dimensionfull quantity is  the temperature $T$. Then, for a quasi-particle with energy $w$ and momentum $q$ moving in the medium there are just 
two independent dimensionless quantities, usually taken as: $ {\mathfrak  q} \,=\,  q/T $ and 
$ {\mathfrak w } \, =\, w/T$. In particular, the authors of  \cite{Kovtun:2005ev} show tables with the five lowest solutions for quasi-normal modes for different channels (field components) where one clearly sees that the general form of the real and imaginary parts of the mode solutions, for a fixed value of $ {\mathfrak  q} $ are of the form: 
\begin{equation}
 \Re ({\mathfrak w }) \equiv  \frac{w_R }{T}  \, =\, C_1 \,;\, \Im  ({\mathfrak w }) \equiv  \frac{ w_I }{T }\, =\, C_2 \,,
 \end{equation} 
where $C_1 $ and $C_2$ are constants that depend on the channel and on the order of the mode. This means that  the real and imaginary parts of the frequency $w$ are just proportional to the temperature. 
At zero temperature, they vanish, since there are no mass parameters in the theory. 

The situation is quite different in AdS/QCD models, where one introduces mass parameters, breaking conformal invariance. In these holographic models there is a non trivial dependence of the real and imaginary parts  of the quasi-normal frequencies with the temperature.  In particular,  the real part $w_R$ is related to the mass of the corresponding quasi-state and assumes, in the limit  $T\to 0 $,  the value of the physical mass of the corresponding hadronic  state in the vacuum. In the present case, the holographic model involves three energy parameters, that make it possible to fit the zero temperature masses and decay constants.  The quasi-normal model solutions are rather non trivial and represent, as we will see in the sequence, the thermal behavior of the heavy vector mesons in the plasma. In order to have a deeper understanding about the effect of the introduction of the background $\phi(z)$ of  eq. (\ref{dilatonModi}) we will discuss the hydrodynamic limit $  {\mathfrak  q} << 1 $ , 
$ {\mathfrak w } << 1 $ in the appendix.

We are going to obtain the quasinormal modes for electromagnetic perturbations by solving the equations of motion using  numerical methods. The shooting method of refs. \cite{Kaminski:2008ai,Kaminski:2009ce,Amado:2009ts,Kaminski:2009dh,Janiszewski:2015ura} is of particular interest for the present work since it is suited to find quasinormal modes of space-times that are only know numerically. The method consist in specify two boundary conditions at the horizon, and then adjust the free parameter given by the frequency. In our case we need solve the equations (\ref{eqTransx}) and (\ref{eqLogx})
\begin{equation}\label{eqTransx2}
\partial_{z}^{2}\mathcal{E}_{\alpha}+\left(\frac{\partial_{z}f}{f}-\frac{1}{z}-\partial_{z}\phi \right)\partial_{z}\mathcal{E}_{\alpha}+\frac{\omega^2-q^2f}{f^2}\mathcal{E}_{\alpha}=0 \, , \,\,\,\,\,\,\,\,\,\,\ (\alpha=1,2)
\end{equation}

\begin{figure}[h]
\label{g67}
\begin{center}
\includegraphics[scale=0.5]{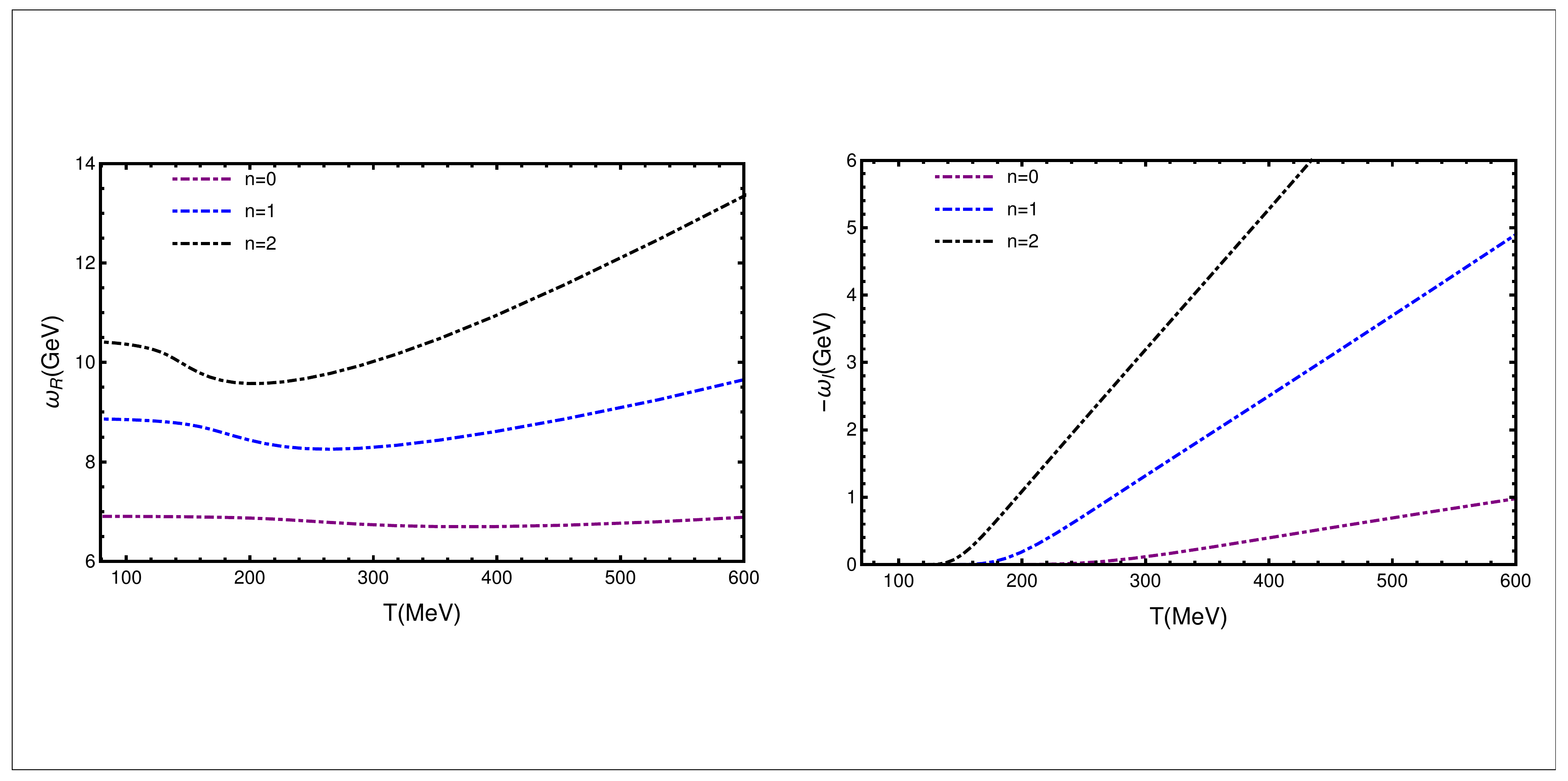}
\end{center}
\caption{Real and imaginary parts of the  frequencies of bottomonium quasinormal modes as function of the temperature. The real part of the frequencies is shown on the left side, and the imaginary part on the right side. }
\end{figure}

\begin{equation}\label{eqLogx2}
\partial_{z}^{2}\mathcal{E}_{3}+\left(\frac{\omega^2}{\omega^2-q^2f}\frac{\partial_z f}{f}-\frac{1}{z}-\partial_{z}\phi \right)\partial_{z}\mathcal{E}+\frac{\omega^2-q^2f}{f^2}\mathcal{E}_{3}=0, \,\,\,\,\,\,\,\,\,\,\ 
\end{equation} 
using the boundary conditions given by the  infalling boundary conditions at the horizon position, presented  in  eq. (\ref{boundary}),
\begin{eqnarray}\label{infa1}
\lim_{z\rightarrow z_h}\mathcal{E}_{j}(z,p)&=& \left(1-\frac{z}{z_h}\right)^{-i\omega/4 \pi T}\,\left[1+a_{1}\left(1-\frac{z}{z_h}\right)+\cdots\right].
\end{eqnarray}
The  second boundary condition is the derivative of the infalling  condition : 
\begin{eqnarray}\label{infa2}
 \lim_{z\rightarrow z_h} \partial_{z} \mathcal{E}_{j}(z,p)&=& \left(1-\frac{z}{z_h}\right)^{-i\omega/4 \pi T}\,\left[\frac{-a_{1}}{z_{h}}+\frac{-a_{2}}{z_{h}}\left( 1- \frac{z}{z_h}\right)\cdots\right] \cr  &-& \frac{i\omega}{(4 \pi T)}\left(1-\frac{z}{z_h}\right)^{-1}\lim_{z\rightarrow z_h}\mathcal{E}_{j}(z,p).
\end{eqnarray}
The coefficients that appear in the infalling condition can be determined inserting this condition into the equations of motions. Similarly to normal modes, the quasinormal modes exist only for a discrete set of frequencies $\omega_{n}$. 

In addition, it is necessary  to  ensure  that  the solution found by solving eqs. (\ref{eqTransx2}) and (\ref{eqLogx2}), using the boundary conditions at the horizon, is a quasinormal mode varying the frequency $\omega$ until the field satisfy the Dirichlet conditions  at the boundary.

It is important to remark that in order to overcome the limitation of the shooting method hen large imaginary part $\omega_{I}$ are present,  it is necessary computing more coefficients of the near horizon expansion (\ref{infa1}) for higher temperatures in order to  find the quasinormal modes. This issue is discussed in \cite{Kaminski:2009ce}. 

\subsection{Quasi-particles at rest in the medium} 
Using the shooting method the quasinormal frequencies were determined as a function of the temperature, for the case of zero momentum $q=0$. 

\begin{figure}[h]
\label{g67}
\begin{center}
\includegraphics[scale=0.5]{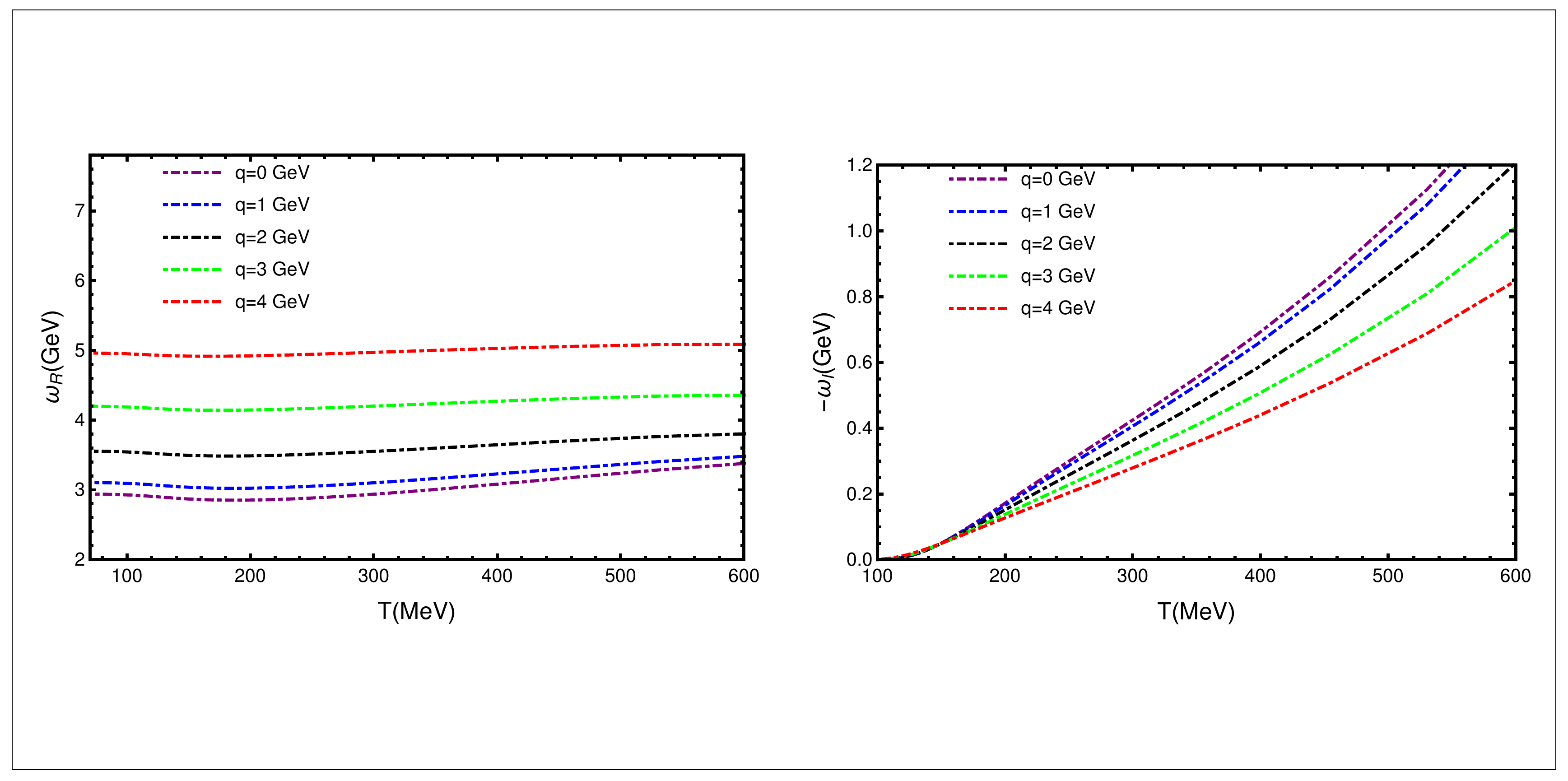}
\end{center}
\caption{Numerical results for the quasinormal modes of the charmonium in the longitudinal direction as function of the temperature. }
\end{figure}
\begin{figure}[h]
\label{g67}
\begin{center}
\includegraphics[scale=0.5]{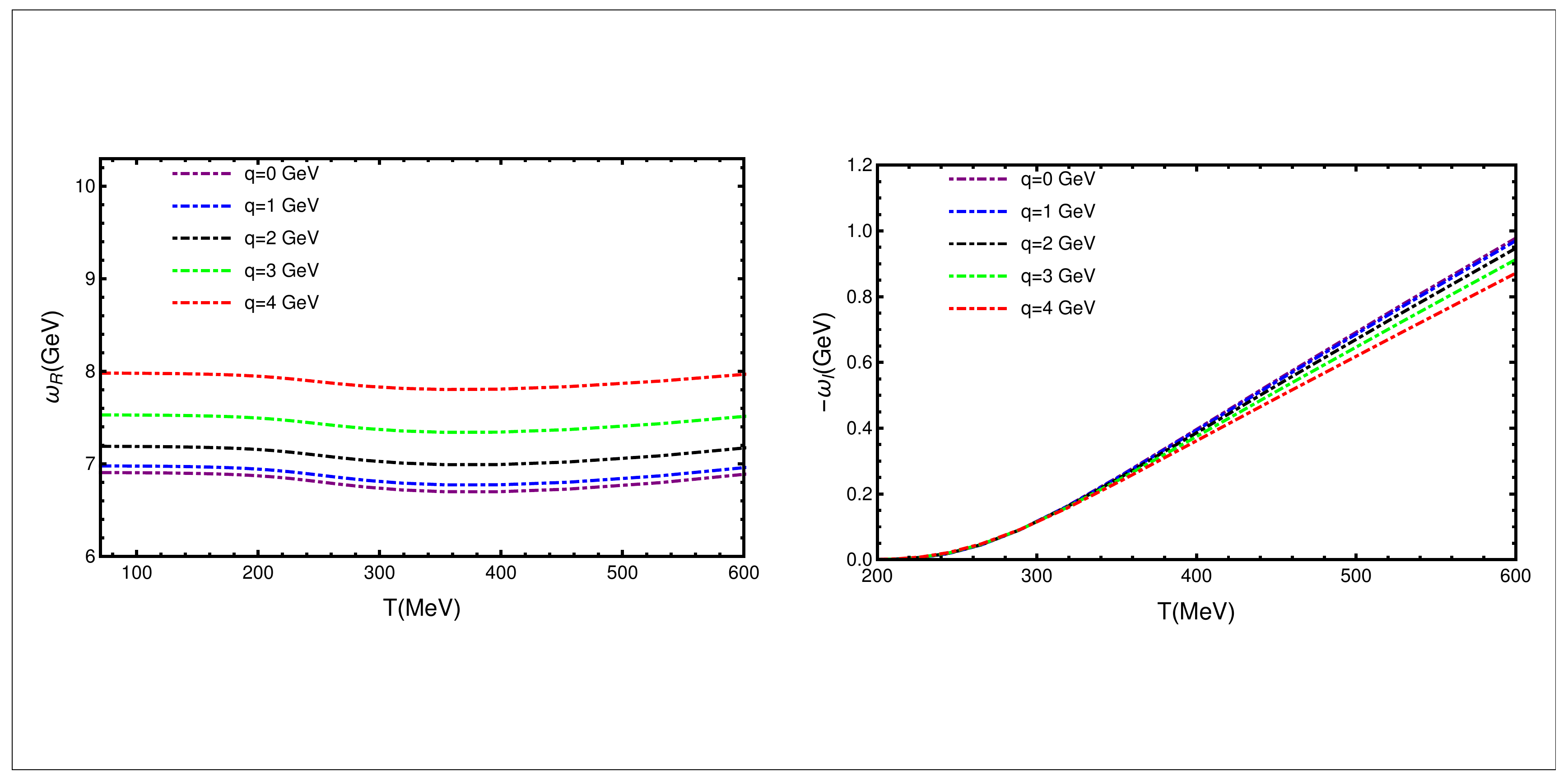}
\end{center}
\caption{Numerical results for the quasinormal modes of the bottomonium in the longitudinal direction as function of the temperature. }
\end{figure}

We show in figures 5 and 6 the results for the real  and imaginary parts  of the frequencies for the first three modes $n=0,1,2$ with $q=0$  for charmonium and bottomonium, respectively. Note that the imaginary part of the frequency is negative and we show in our plots the values multiplied by minus one. From these figures one notes that in the region of high temperatures, the frequencies show a linear behaviour that is in agreement with ref.\cite{Kovtun:2005ev}. 

On the other hand, one can note that in the zero temperature limit, the real part of the quasinormal frequencies coincide with the corresponding mass spectrum of Table 1 and 2 and the  behaviour at low temperature is not linear as in the  high temperature. One also notes that the width for the charmonium grow  faster than for bottomonium, showing that the dissociation process is more intense for charmonium. 

\subsection{Quasi-Particles in motion: dispersion relations}

\subsection*{Longitudinal Perturbations}

 We found  the quasinormal frequencies  with finite momentum in the longitudinal sector using the shooting method in the equation (\ref{eqLogx2}).  The variations of the quasinormal modes for different momentum as a function of the temperature for the ground state for charmonium and bottomonium are shown in Figs 7 and  8 respectively. As one can see, the real part of the frequencies increases with $q$. Such behaviour is consistent with the results in figures 3 and 4, where the position of the peaks of the spectral function shifts to high energies when the value of  the momentum increases. A different thing happens with the imaginary part of the frequencies (width of the peaks), which decreases with $q$ as is shown in the right panel of the figures 7 and 8.

\subsection*{Transverse Perturbations}

The results of the quasinormal modes for the transverse perturbations are presented in  figures 9 and 10 as a function of the temperature. On one hand, the real part of the quasinormal frequencies have a similar behaviour  to that of the longitudinal sector. On the other hand, the imaginary part of $\omega$ increases with momentum, in contrast to the longitudinal case. 
These results indicate that dissociation effect increases  due to motion relative to the medium in the direction transverse to the polarization of the heavy vector mesons.

\begin{figure}[h]
\label{g67}
\begin{center}
\includegraphics[scale=0.5]{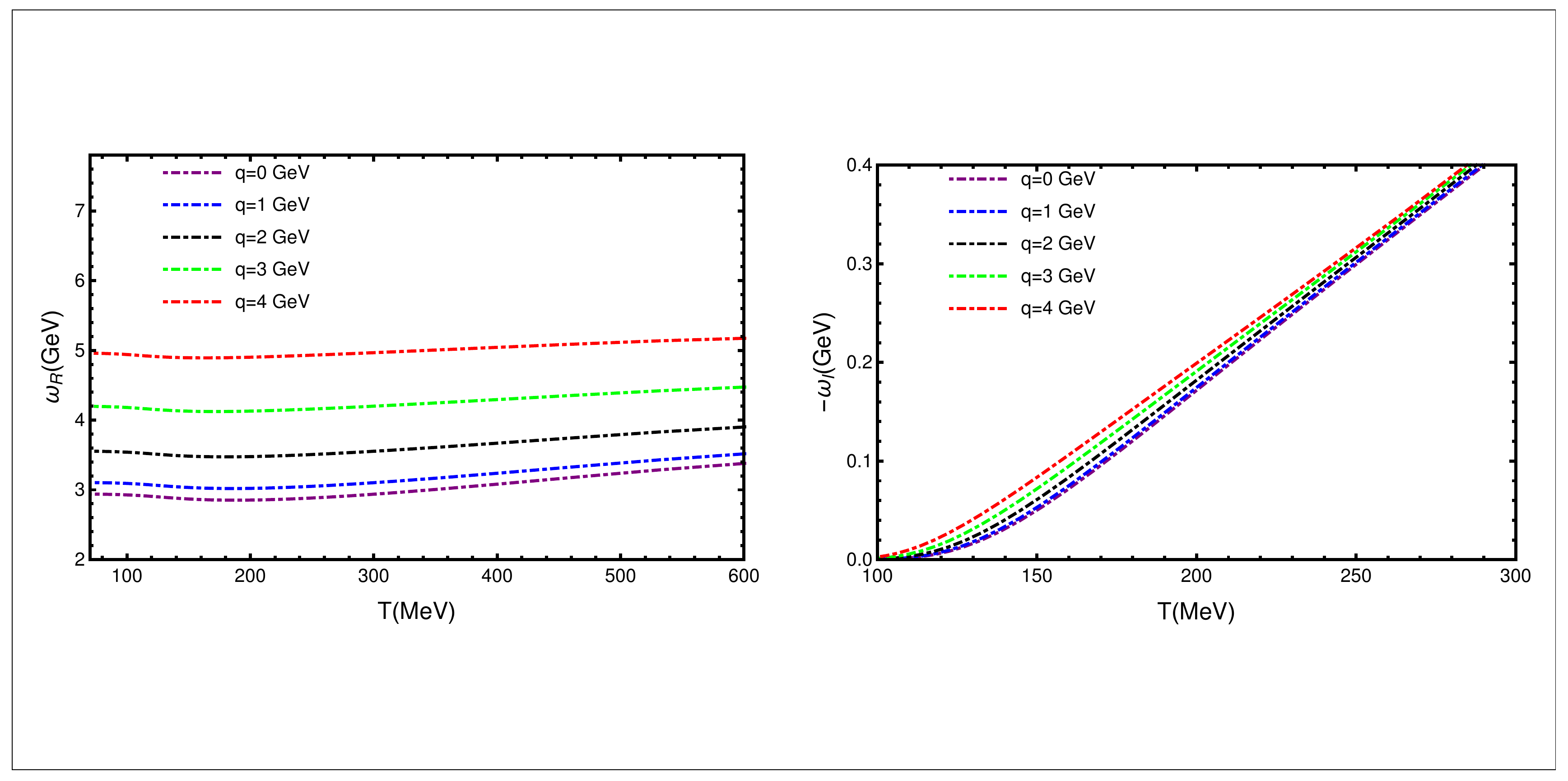}
\end{center}
\caption{Numerical results for the quasinormal modes of the charmonium in  the transverse direction as function of the temperature.}
\end{figure}
\begin{figure}[h]
\label{g67}
\begin{center}
\includegraphics[scale=0.5]{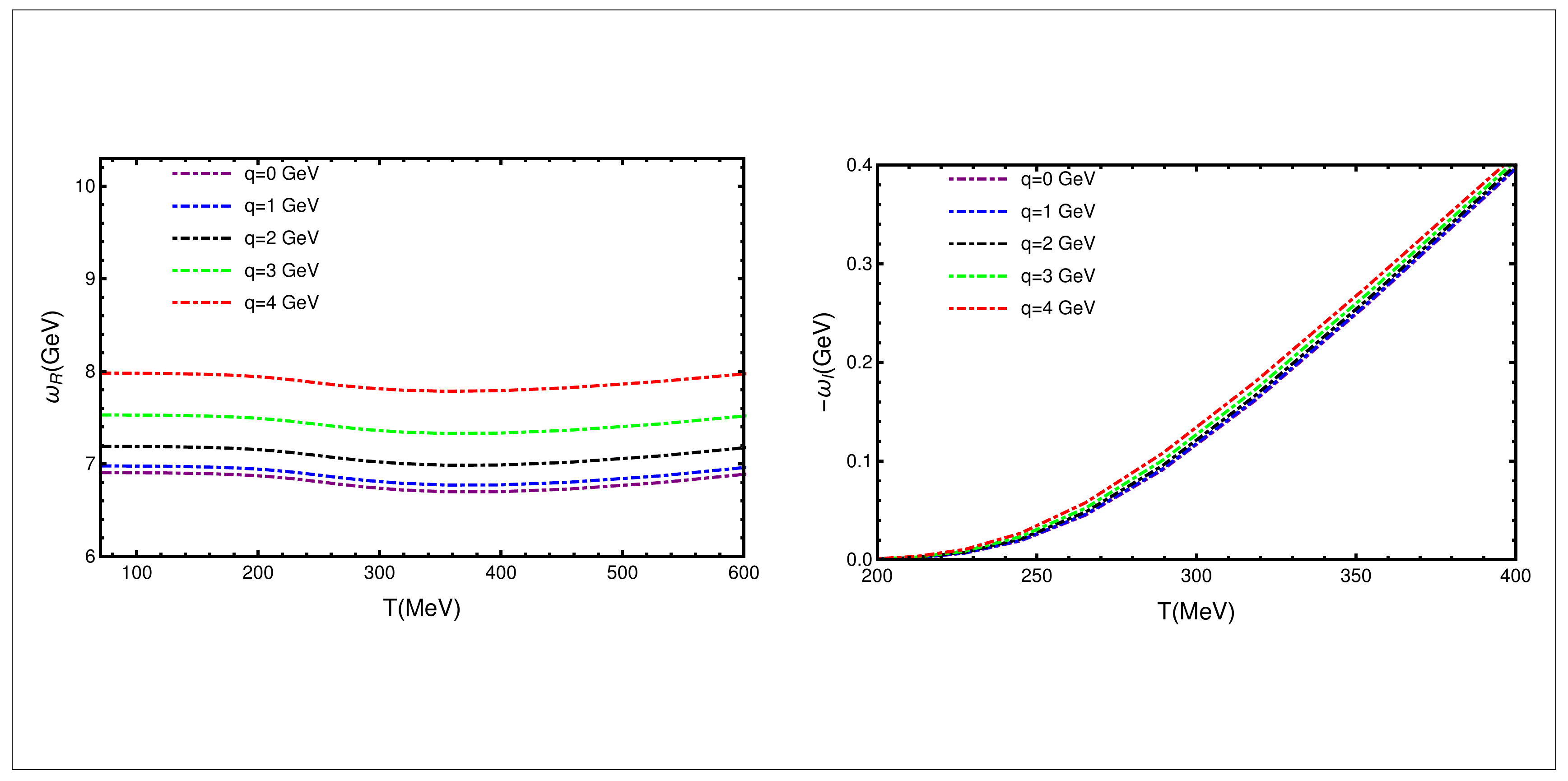}
\end{center}
\caption{Numerical results for the quasinormal modes of the bottomonium in  the transverse direction as function of the temperature.}
\end{figure}

\section{Discussions of the results and Conclusions} 
Using a holographic bottom up model, we have studied the real and imaginary parts of the  quasinormal modes frequencies corresponding to charmonium and bottomonium states. 
The dependence with the temperature for quasi-particles at rest, or moving with respect to the medium,  was investigated. 
For comparison, the spectral function for these heavy mesons was also obtained.  

The outcomes from the quasinormal modes are consistent with the spectral function behaviour, as 
we now discuss.  
For mesons at rest, the real part of the frequency shows a tendency to increase with the temperature.
In terms of the spectral function this is translated into the increase in the position of the location of the peak. The imaginary part of the frequencies monotonically increase, a behaviour that is translated into the decrease in the height and an increase in the width of the peaks.
 
 For mesons in motion with respect to the medium, the real part of the quasinormal modes frequencies increase with the value of the momentum, for both longitudinal and transverse motion. This is translated in terms of the spectral function into the  increase in the value of the frequency where the peaks are  located. 
 For the imaginary part of the frequency, the behaviour is different if the motion is in the direction of the polarization (longitudinal case) or transverse to the polarization. 
 In the longitudinal case the imaginary part decreases with the momentum. This behaviour is consistently reproduced in the spectral function as an increase in the height of the peaks.
 For transverse motion,  the behaviour is the opposite. The imaginary part of the frequency increases with the momentum and, correspondingly, the height of the peaks of the spectral function decrease.

The dissociation of quarkonium states inside the quark gluon plasma was studied before using 
lattice QCD  \cite{Umeda:2000ym,Asakawa:2003re,Datta:2003ww,Jakovac:2006sf,Aarts:2007pk,Rothkopf:2011db,Aarts:2011sm,Aarts:2010ek,Aarts:2012ka,Aarts:2013kaa,Karsch:2012na},
 QCD sum rules \cite{Morita:2007pt,Morita:2007hv,Song:2008bd,Morita:2009qk,Gubler:2011ua,Suzuki:2012ze}, other previous holographic models \cite{Kim:2007rt,Fujita:2009wc,Noronha:2009da,Kaminski:2009ce}, 
 effective field theories \cite{Brambilla:2008cx,Digal:2005ht}  and potential models  \cite{Alberico:2005xw, Mocsy:2007yj,Mocsy:2007jz,Petreczky:2010tk}. There are also experimental investigations as in ref. \cite{Adare:2014hje}.  In particular, in \cite{Adare:2014hje} one finds a plot, on Figure 1, with a compilation of results for quarkonium dissociation temperatures from those different approaches.    The dissociation for charmonium at zero chemical potential happens for the ratio $ T/T_c$ in the range of 1.5 to 3.0. For bottomonium, the dissociation occurs in a range, that varies with the model,  from $2  \le  T/T_c \le 4 $. 
 
 The effective holographic model presented here describes heavy mesons. That means: mesons made with quark flavors $b$ and $c$.  On the other hand, the quark gluon plasma is formed by the dissociation of the light quarks. A reasonable estimate for the critical temperature here is to use the value obtained in holographic soft wall model for the (light) rho meson\cite{BallonBayona:2007vp}: $ T_c \sim 190$ MeV. 
Then, the dissociation temperatures found here are consistent with the other predictions. 

Regarding the effect of the motion relative to the plasma frame, reference \cite{Kaminski:2009ce} analysed mesons in motion in a plasma using the D7 brane model. An increase in the real part of the quasi-normal frequencies with the value of the momentum was found for both transverse and longitudinal motion.  The same kind of result that we found here. 
For the imaginary part of the frequency, the authors of \cite{Kaminski:2009ce} found a decrease in the absolute value with the momentum for the longitudinal case,  the same qualitative behavior found here. For the transverse case, the behavior that they found for the imaginary part of the frequencies depends on a parameter that characterizes the D7 brane embedding.

\section{Appendix} 

We can understand better the effect of the introduction of the  background $\phi(z)$ of  eq. (\ref{dilatonModi}) in the holographic model by considering the hydrodynamical limit $ {\mathfrak q} << 1 $ , 
$ {\mathfrak w } << 1 $.
One considers the charge diffusion hydrodynamic dispersion relation, corresponding to the lowest quasinormal frequency \cite{Policastro:2002se}. Using the equation of motions  for the vector field, together with appropriate boundary conditions give the dispersion law  $\omega(q)=-Dq^2$ \cite{Policastro:2002se}.
The charge diffusion constant can be obtained following  the  membrane paradigm as in ref. \cite{Kovtun:2003wp} 
\begin{equation}\label{diffusion}
D=\frac{\sqrt{-g(z_h)}}{g_{eff}^{2}(z_h)\sqrt{-g_{tt}(zh)g_{zz}(z_h)}}\int^{z_h}_{0}dz\frac{-g_{tt}(z)g_{zz}(z)}{\sqrt{-g(z)}}g_{eff}^{2}(z),
\end{equation} 
where  $g_{eff}$ is a $z$ dependent coupling. Let's compute  the diffusion constant for the AdS/CFT case, that corresponds to the  strongly coupled  $\mathcal{N}=4$ SYM plasma. The metric  of the gravity dual is given by  eq. (\ref{metric2}) and $g_{eff}^2=g^2_{5}$. Then, one finds using  (\ref{diffusion})
\begin{equation}\label{diffusionADS}
D_{\mathcal{N}=4 \, SYM}=\frac{1}{2\pi T}.
\end{equation}
where the  only parameter is the temperature. This is the result of refs. \cite{Policastro:2002se,Kovtun:2003wp}.

 In the case of the holographic model considered in this article the metric is the same but there is a $z$ dependent coupling given by $g_{eff}^2=e^{\phi(z)}g^2_{5}$. Therefore, the diffusion constant can be written as
 \begin{equation}\label{diffusionSoftwall}
D=\frac{e^{-\phi(z_h)}}{z_h}\int^{z_h}_{0}dzze^{\phi(z)}.
\end{equation}  
For the dilaton profile (\ref{dilatonModi}) that includes a hyperbolic tangent term we could not find an analytic solution for $D$. The numerical solutions, for  charmonium and  bottomonium,  were calculated and the results for the product $ D 2 \pi T $ are shown in figure (11) as functions of the temperature divided by the mass of the first state.  One can observe that, consistently, in the high temperature (conformal) limit  the diffusion coefficient for both cases approaches the AdS/CFT value: $D = 1/2\pi T$. But for finite temperatures there is a non trivial dependence on $T$. 

On the other hand, if one considers a dilaton profile similar to (\ref{dilatonModi}) but  without the hyperbolic tangent: $\phi(z)=k^2z^2+Mz$, one finds an analytical solution that serves as an illustrative example of the effect of the background scalar field. The solution found using (\ref{diffusion}) is:
\begin{eqnarray}
\label{diffusionSoftwall2}
D= & & e^{-\frac{(2k^2+M\pi T)^2}{4k^2\pi^2T^2}}\pi T \biggl(  2 e^{M^2/4k^2}k\left(-1+e^{\frac{(k^2+M\pi T)}{\pi^2T^2}}\right) + M \sqrt{\pi} \Erfi{\left[\frac{M}{2k}\right]} \cr
&-&  M \sqrt{\pi} \Erfi{\left[\frac{M}{2k}+\frac{k}{\pi T}\right]} \biggl)\,,
\end{eqnarray} 
where $\Erfi(x)=-i\Erf(i x)$ is the imaginary error function. This analytical form shows us, for this simpler background, how the energy  parameters $k$ and $M$ contribute to the charge diffusion coefficient. This expression shows the non trivial dependence of $D$ on the temperature for a holographic model, that contrasts with the trivial dependence of the conformal AdS/CFT case. 

\begin{figure}[h]
\label{g67}
\begin{center}
\includegraphics[scale=0.6]{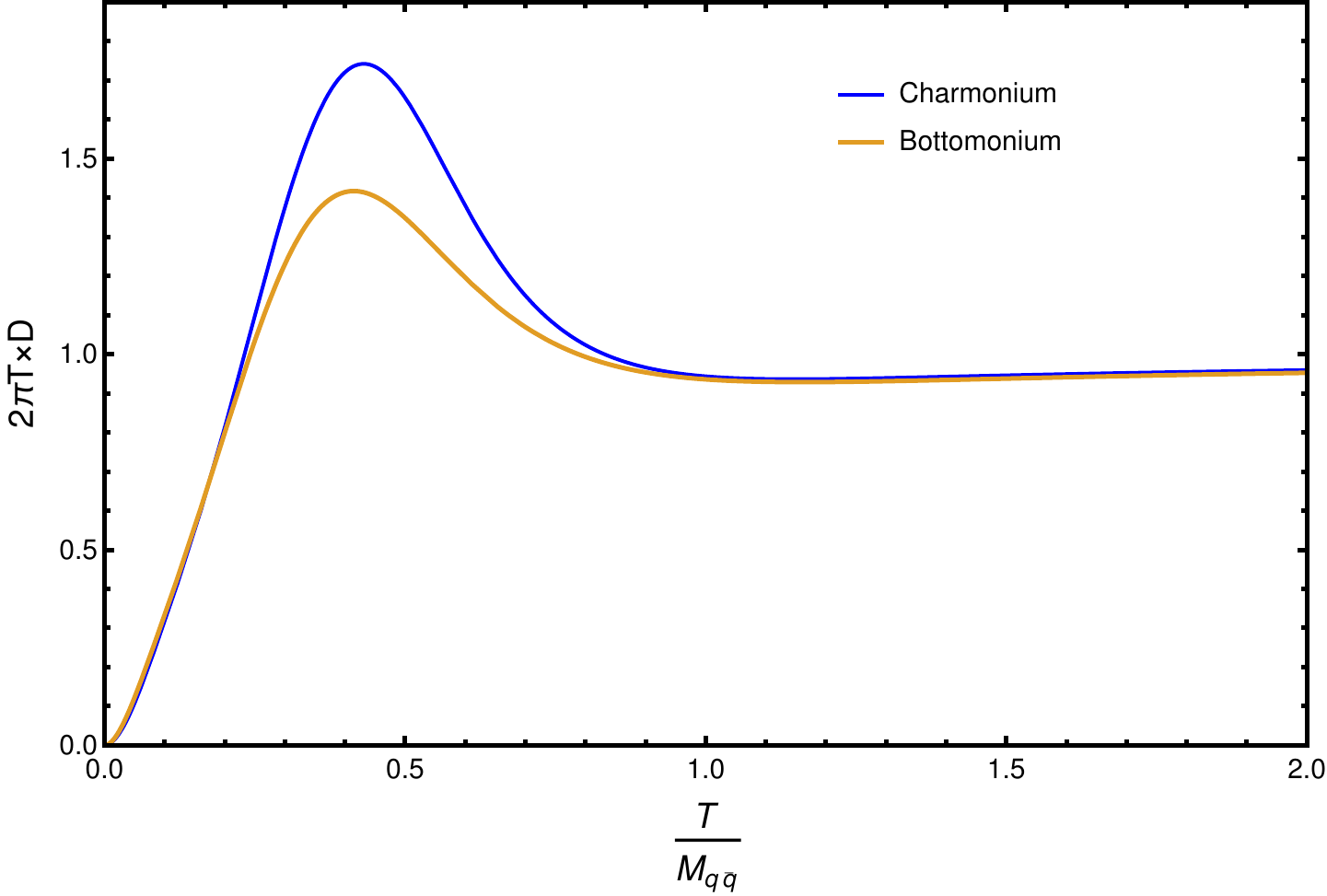}
\end{center}
\caption{Numerical results for the charge diffusion coefficient of  charmonium and bottomonium multiplied by $ 2 \pi T $  as function of the temperature (rescaled by the mass of the first hadronic state). }
\end{figure}

\noindent {\bf Acknowledgments:} N.B. is partially supported by CNPq (Brazil) under Grant No. 307641/2015-5 and L. F. Ferreira is supported by the National Council for Scientific and Technological Development – CNPq.

 \end{document}